\documentstyle[graphicx,twocolumn,aps]{revtex}

\tightenlines

\begin{document}
\draft
\preprint{ND Atomic Theory Preprint 99/04}
\title{
Large Contributions of Negative Energy States to Forbidden Magnetic-Dipole Transition Amplitudes in Alkali-Metal Atoms 
}

\author{I. M. Savukov, A. Derevianko, H. G. Berry, and W. R. Johnson}
\address{Department of Physics, 
University of Notre Dame, Notre Dame, Indiana 46556}
\date{\today}
\maketitle

\begin{abstract}
     {
     The influence of negative-energy states (NES) on forbidden
     magnetic-dipole $ns_{1/2}-(n+1)s_{1/2}$ transitions in alkali-metal atoms is investigated.
     We find that
     the NES contributions are significant in almost all cases and for rubidium 
     reduce the transition rate by a factor of 8. We tabulate magnetic-dipole ($M_1$) transition amplitudes for the alkalis. Our $M_1$ value for cesium, where accurate measurements are available, differs from experiment by 16\%. We briefly discuss the feasibility
     of an experimental test of NES effects. 
}
 
\end{abstract}


It is well known (see, for example, the discussion in Ref.~\cite{Sucher}) that the Dirac-Coulomb Hamiltonian has no bound state eigenfunctions
in the presence of the electron-electron interaction. The  {\em no-pair} Hamiltonian $H_{n.p.}$ 
derived from
quantum electrodynamics has been proposed by Brown and Ravenhall~\cite{BrownRavenhall51} for use in
relativistic atomic calculations. Although $H_{n.p.}$ leads to very accurate energies, the omitted effects of electron-positron pairs  can be significant for the forbidden
magnetic-dipole ($M_1$) transition amplitudes.
The first discussion of pair corrections to the $M_1$ decay rate for the $2\;^3\!S_1$ state in helium was given by
Feinberg and Sucher~\cite{FeinbergSucher71} in 1971. Later, when new lifetime measurements of the $2\,^3\!S_1$ metastable state of heliumlike ions
became available, more accurate calculations of $M_1$ 
were performed by several theoretical groups. 
Lindroth and Solomonson~\cite{LindrothSalomonson90} 
numerically demonstrated the detailed cancellation of one-pair 
diagrams  in the case of heliumlike argon. The decay rates of the same transitions for  
heliumlike ions with $Z=2-100$ were calculated within relativistic configuration-interaction
approach by Johnson, Plante, and Sapirstein~\cite{Johnson95}. They treated contributions of NES using second-order many-body perturbation theory (MBPT).
Indelicato~\cite{Indelicato96} considered the effects of NES for $2\,^3\!S_1$ $M_1$ decay 
in the multiconfiguration Dirac-Fock (MCDF) approach.
Recently we studied NES contributions
to transition amplitudes in more detail ~\cite{Derevianko98}. In the Pauli approximation, we derived
an effective one-pair operator that explicitly reveals
cancellation between Coulomb and Breit two-body diagrams and, by using
it, found a transition without such cancellation: the neutral helium
$2^3S_1-3^3S_1$  transition has a very large NES contribution which reduces
the $M_1$ rate by a factor of 2.9.

 There have been no other
calculations treating NES contributions to $M_1$ transitions systematically
in multi-electron systems, except for He- and Be-like transitions~\cite{Safronova99}. 
In this letter, we report second-order MBPT
calculations of magnetic-dipole $ns_{1/2}-(n+1)s_{1/2}$  transitions in the
alkalis including the analysis of one-pair effects. We have discovered 
an unusually large NES contribution, which reduces the transition rate by 
a factor of 8 in Rb, and propose measurements to test NES effects. The results of 
our calculations for Cs are compared to a previous theoretical determination~\cite{Dzuba85} and to experimental values~\cite{BennettWieman99,Bouchiat88}.
For the other alkalis, no measurements exist.

We can argue that forbidden magnetic-dipole amplitudes are the most
sensitive among electro-magnetic transition amplitudes 
to the accuracy of the relativistic description of an atomic system. As we will demonstrate,
several factors contribute to the result: correlation effects, 
spin-orbit interaction, Breit interaction, retardation effects, and, finally,
the negative-energy contributions.
The relativistic retarded magnetic-dipole matrix element  can 
be represented as~\cite{Johnson95} (atomic units are used throughout the letter)
\begin{eqnarray}
\lefteqn{ 
\langle i||M_{1}||j\rangle =c\;\langle -\kappa _{i}||C_{1}||\kappa
_{j}\rangle \left( \kappa _{i}+\kappa _{j}\right) 
 \times } \nonumber \\ &
\int_{0}^{\infty } &\frac{3}{%
k}\,j_{1}\left( kr\right) \,\,\left( G_{i}F_{j}+F_{i}G_{j}\right) \;dr \, .
\label{EqnFullM1}
\end{eqnarray}
Here $G$ and $F$ are the radial parts of large and small components of
atomic orbitals, $k$ is the photon wavenumber, and $\kappa = \left(
j+\frac{1}{2} \right) (-1)^{j+l+1/2}$. 
In the long-wavelength limit and Pauli approximation this relativistic
expression reduces to a conventional non-relativistic operator
\begin{equation}
M_{1}= L+2S \,.  \label{EqPauliM1} 
\end{equation}
Even if the general angular selection rules for the $M_{1}$ operator are
satisfied, this matrix element vanishes when the radial wavefunctions are
orthogonal, i.e. if $\kappa _{i}=\kappa _{j}$, but $n_{i}\neq n_{j}$.
The Einstein $A$-coefficient for the $M_1$ transition 
$|I\rangle \rightarrow |F\rangle$ 
is expressed in terms of the reduced matrix element as
\[
A_{M1} = \frac{k^3}{3 c^2}  \frac{|\langle F||M_{1}||I\rangle|^2}{2J_I +1}   
  \, .
\]

\begin{figure}
  \centerline{\includegraphics*[scale=0.75]{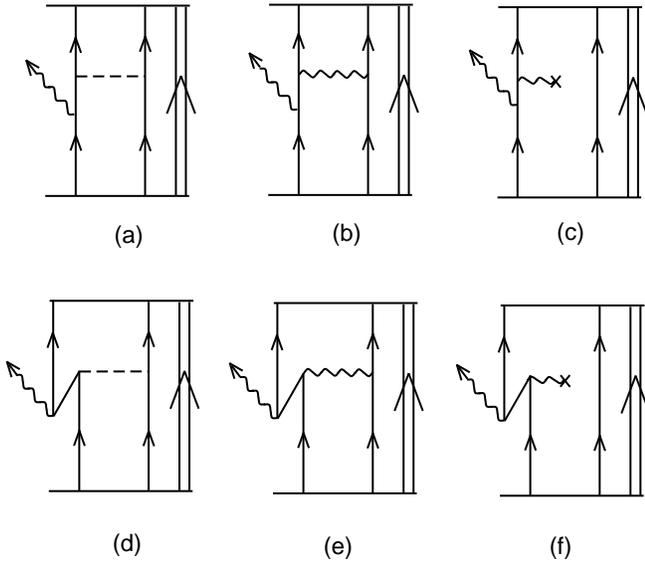}}
  \caption{  
  Principal Feynman-like time-ordered diagrams contributing to $M_1$ amplitude 
  in the second-order. The wavy lines represent photons and solid lines represent
  atomic electrons. Double vertical solid line designates inactive (observing) electrons.
  Diagrams (a) and (d) are due to Coulomb interaction, (b) and (e) due to two-body 
  static Breit interaction, and (c) and (f) due to one-body static Breit interaction.
  Upper panel of diagrams represents {\em no-pair} contributions, and the lower
  panel --- contributions from negative-energy states.    
 \label{Fig-feynman}} 
\end{figure}

We start our consideration by utilizing second-order MBPT built on the 
"frozen-core" Dirac-Hartree-Fock (DHF) potential. This approximation includes both leading correlation and NES effects.
We consider matrix elements of the magnetic-dipole operator $z$ between two valence states $v$ and $w$. For
the purposes of this paper, the valence state $v$ represents the ground
state orbital $ns_{1/2}$ and the state $w$ represents the first excited $s$ state 
$(n+1)s_{1/2}$. 
The first order value is given by the matrix element taken
between DHF orbitals $z_{wv}$. The second-order correction 
including both Coulomb ($g_{ijkl}$) and Breit ($b_{ijkl}, b_{ij}$) 
interactions is given by    
\begin{eqnarray}
Z_{wv}^{(2)} &=&\sum\limits_{i\neq v}\frac{z_{wi\,}b_{iv}}{\epsilon
_{v}-\epsilon _{i}}+\sum\limits_{i\neq w}\frac{b_{wi}\,z_{iv}}{\epsilon
_{w}-\epsilon _{i}}+  \nonumber\\
+\sum\limits_{na} && \! \! \frac{z_{an}(\widetilde{g}_{wnva}+\widetilde{b}_{wnva})}{%
\epsilon _{a}+\epsilon _{v}-\epsilon _{n}-\epsilon _{w}}+\sum\limits_{na}%
\frac{(\widetilde{g}_{wavn}+\widetilde{b}_{wnva})\,z_{na}}{\epsilon
_{a}+\epsilon _{w}-\epsilon _{n}-\epsilon _{v}} \, . \label{EqnZ2MBPT}
\end{eqnarray}
Here antisymmetrized matrix elements are defined as $\widetilde{g}_{ijkl} = {g}_{ijkl} - {g}_{ijlk}$.
The one-body matrix element of the Breit interaction is $b_{ij}= \sum_a \widetilde{b}_{iaja}$. 
In the above expressions, index $a$ runs over core orbitals and $n$ extends over
virtual orbitals. It is important to note that virtual states include both
excited positive and negative-energy orbitals. 
Corresponding Feynman diagrams are given in Fig.~\ref{Fig-feynman}.
We emphasize that in the {\em no-pair} approximation
the summation over virtual states would be limited only
to the positive-energy states, i.e. include only the diagrams in the upper panel of
Fig.~\ref{Fig-feynman}.
The inclusion of the Breit interaction in our analysis is crucial, because
its contribution to the matrix element is of the same order as for the Coulomb 
interaction. We use a static limit of the Breit interaction, since
the energy of the transverse photon is determined by the energy 
difference of the real electrons, not by that of the virtual electron. 
The numerical calculations were performed with a relativistic B-spline basis set 
representation~\cite{splines} obtained in a cavity of radius 40 atomic units 
and included 40 positive- and 40 negative-energy wavefunctions for each partial wave.

The results of our calculations
are presented in  Table~\ref{TabSum}.  Note that the first-order Dirac-Hartree values 
dominate for lighter
atoms and become less significant for cesium and francium. 
This is due to larger second-order {\em no-pair}
contributions. The values of NES contributions (in the third row) appear 
to be roughly proportional
to $Z$. In the case of cesium they constitute only a small fraction 4\% which 
is even smaller for francium 0.6\% due to large {\em no-pair} second-order contributions. 
However, the NES fraction reaches 19\% in potassium. 
The rubidium case is the most surprising: there is
cancellation of the two {\em no-pair} terms and consequently a strong dependence of the total 
value on the negative-energy corrections. Although such cancellation in second order 
may be coincidental, and more accurate calculations may be necessary,
we conclude that a measurement of the $M_1$ transition in rubidium could
provide an excellent test of the NES contributions.

The large relative contribution of NES for forbidden magnetic-dipole
transitions is caused by several factors. First, due to unique properties
of the $M_1$ operator~(\ref{EqPauliM1}), the {\em no-pair} amplitude is severely suppressed 
(by a factor of $\alpha^2$ in the lowest order).
Second, the magnetic-dipole operator $(M_1)_{ij}$ in Eq.~(\ref{EqnFullM1}) 
contains an integral of large and small components
of Dirac wavefunctions. For positive-energy states the small component is
significantly weaker than the large component (by a factor of $\alpha Z$ 
for hydrogen-like ions). 
For NES, the situation
is reversed, and the small component is much larger. 
In addition, the Pauli approximation 
expression~(\ref{EqPauliM1}) with its particular $\delta$-function-like
properties is no longer valid and one obtains much larger values for $M_1$ matrix elements
between negative and positive states, than from positive-positive matrix elements.  
Finally, the energy denominators of order $2mc^2$ bring the NES contributions
to the same level  as the contribution from the ``regular'' positive-energy states.      
As seen from Table~\ref{TabNeg}, NES contributions from the Breit interaction are comparable to those from the Coulomb diagrams because the Breit operator mixes large and small components.

We note that for Rb, Cs and Fr, correlation effects are very important leading
to contributions larger than the lowest-order DHF values. 
The mechanism has been discussed by Dzuba {\em et al.}~\cite{Dzuba85}.  
In the
Pauli approximation, the $M_1$ matrix element is proportional to the 
integral of the product of  the large
components between the states involved.  
In the first-order forbidden transitions  
$ns_{1/2} - (n+1)s_{1/2}$ between
states with different principal quantum numbers, the radial wavefunctions are orthogonal and the $M_1$
rate is zero. Although it is not zero beyond Pauli approximation, it is strongly suppressed.
The situation is quite different for $p_{1/2}$ and $p_{3/2}$ matrix elements which are non-zero due
to overlapping radial wavefunctions caused by the spin-orbit interaction. As a result, the second
order contributions dominate due to such matrix elements connecting core and exciting states. 
This correlation effect becomes overwhelming for heavier elements where spin-orbit coupling is important.

In Table~\ref{TabExp-Th}, we compare our cesium results for the magnetic-dipole reduced matrix element  
with calculations of Dzuba {\em et al.}~\cite{Dzuba85} and with  measurements from several 
experimental groups. The transition amplitude used in~\cite{Dzuba85} is
related to the reduced matrix element expressed in atomic units as
\[ 
(M_1)_{\mathrm{ampl.}}   = \frac{1}{\sqrt{6} } \;
\langle n_{w}s_{1/2}||M_{1}||n_{v}s_{1/2}\rangle \times |\frac{\mu_{\rm B}}{c}|    
\, .
\]
The experimental entries for the $M_1$ matrix element in Table~\ref{TabExp-Th} were obtained 
from measurements of $M_1^{\rm hf}/M_1$ and a semiempirical value~\cite{Bouchiat88} 
of the off-diagonal hyperfine mixing amplitude $M_1^{\rm hf} = 0.8094(20)\times 10^{-5} |\mu_{\rm B}/c|$.  
The result of our
work, despite approximate treatment of correlation effects, 
is in reasonable  agreement (16\%) with the experimental results.
Since the negative-energy effects are marginally smaller
than these deviations, it is not possible to draw definitive conclusions about NES effects from
available experiments in cesium. 
The second-order expression~(\ref{EqnZ2MBPT}) is a leading term of the random-phase approximation (RPA).
The  calculations of Dzuba {\em et al.}~\cite{Dzuba85} implicitly included the
effect of negative-energy states due to the reduction of RPA-like diagrams
to the form of a differential equation. However, their analysis did not take
into account the Breit interaction. Such an approach misses an important negative-energy
contribution. Indeed, we demonstrate in Table~\ref{TabNeg} that
the NES contribution from the Breit interaction is much larger than
that arising from the Coulomb interaction.

\begin{figure}
  \centerline{\includegraphics*[scale=0.6]{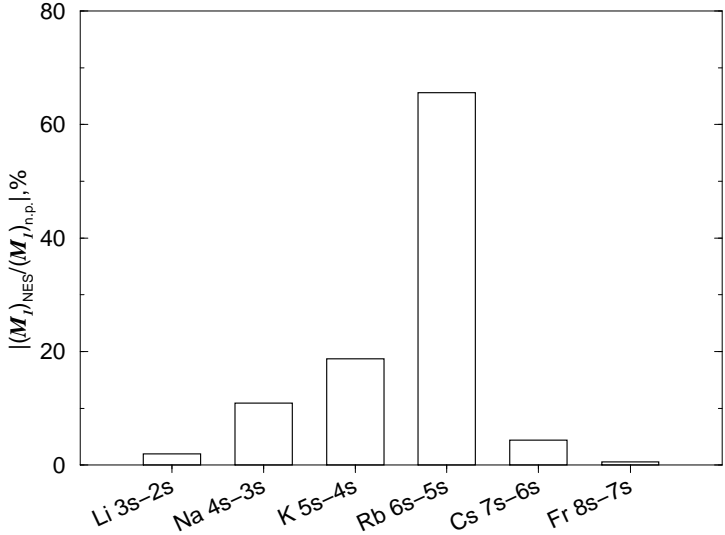}}
  \caption{ 
            The relative contributions to the magnetic dipole ($M_1$) 
            matrix element $ns - (n+1)s$ in
            alkali atoms: the ratio of the NES contributions (row 3 of Table~\protect \ref{TabSum}) 
            to the total {\em no-pair}
            contributions (sum of rows 1 and 2 of Table~\protect \ref{TabSum} ).
 \label{Fig-relNES}} 
\end{figure}

The theoretical calculations of the $M_1$ transition amplitudes in the alkalis clearly demonstrate the important
role of negative-energy states. We now discuss experimental possibilities to test these
contributions.  We compare the NES fractional contributions, defined as the ratio of NES to {\em no-pair}
contributions,  in different alkali-metal atoms in Fig.~\ref{Fig-relNES}.  
In the light alkalis (Li, Na, K) the effect is proportional to $Z$ and is maximal for K
(19\%).  For heavy atoms such as Cs and Fr, it is small because of large {\em no-pair} contributions. 
Rubidium, in the middle, has a very large relative effect (65\%) and is the most promising. If
measurements in the  other alkalis reach the precision achieved in Cs, then all alkalis except Fr will be good
candidates for testing NES contributions. 
The accuracy of the calculations, on the other hand, can impose
even more severe restrictions than experiment. The accuracy of our calculations, as seen in the
deviation from the experiment for Cs, is about 16\%; it is expected to 
be better for lighter elements.
More accurate (1\%) {\em no-pair} calculations are possible, for example, 
in the relativistic single-double approximation~\cite{Safronova98}. For Li, precise {\em no-pair}
configuration-interaction calculations~\cite{KTCheng-LiCI95} are also possible with an accuracy much better
than 1\%.

In conclusion, we have presented the results of the second-order MBPT calculations for the forbidden
$M_1$ transitions in the alkalis.  Comparisons with experimental and theoretical data for cesium have
been made.  We have found very large negative-energy state contributions to the $M_1$ transition
amplitudes in the alkalis.  The NES amplitude is dominant in the case of rubidium, which could
provide the best experimental test of negative-energy contributions in atomic structure.  

We thank Savely Karshenboim for useful discussions and Eugene Livingston for his comments
on the manuscript.
The work of AD and WRJ was supported in part by 
NSF Grant No.\ PHY 99-70666.

\begin{table}[tbp]
\caption{ Contributions to reduced matrix elements of the $M_1$
operator in atomic units multiplied by a factor of $10^{5}$.
Row 1, lowest-order DHF value; row 2, second-order 
{\em no-pair} contribution; row 3, negative-energy state contributions in second
order; and row 4, total value of $M_1$ matrix element. }
\label{TabSum}
\begin{center}
\begin{tabular}{lrrrrrr@{}l}
\multicolumn{1}{c}{} & 
\multicolumn{1}{c}{Li} & 
\multicolumn{1}{c}{Na} &
\multicolumn{1}{c}{K} &
\multicolumn{1}{c}{Rb} &
\multicolumn{1}{c}{Cs} &
\multicolumn{2}{c}{Fr} \\
\multicolumn{1}{c}{$Z$} & 
\multicolumn{1}{c}{3} & 
\multicolumn{1}{c}{11} &
\multicolumn{1}{c}{19} &
\multicolumn{1}{c}{37} &
\multicolumn{1}{c}{55} &
\multicolumn{2}{c}{87} \\
\hline
I & 0.91 & 1.16 & 1.15 & 1.38 & 1.51                     &    2&.09 \\ 
II, {\em no-pair} & 0.12 & 0.03 & -0.08 & -1.86 & -10.69 & -116& \\ 
II, NES & 0.02 & 0.13 & 0.20 & 0.31 & 0.40               &    0&.64 \\ 
Total & 1.05 & 1.06 & 1.27 & -0.17 & -8.78               &  -113&  
\end{tabular}
\end{center}
\end{table}

\begin{table}
\caption{Breakdown of negative-energy state contributions to the reduced
$M_1$ matrix element in atomic units, multiplied by a factor $10^{5}$.
Column ``Coulomb'' represents contributions from diagram (d),
column ``Breit two-body'' --- from diagram (e), and column ``Breit one-body'' --- from diagram (f)
of Fig.~\protect\ref{Fig-feynman}.
}
\label{TabNeg}
\begin{center}
\begin{tabular}{lrrrr}
\multicolumn{1}{c}{} & \multicolumn{1}{c}{Coulomb} & 
\multicolumn{2}{c}{Breit} & 
\multicolumn{1}{c}{Total} \\ 
\multicolumn{2}{c}{} & 
\multicolumn{1}{c}{two-body} & \multicolumn{1}{c}{one-body} &
\multicolumn{1}{c}{}\\
\hline
Li & -0.015 & 0.067 &-0.029 & 0.024 \\ 
Na & -0.020 & 0.106 & 0.047 & 0.133 \\ 
K  & -0.022 & 0.112 & 0.106 & 0.197 \\ 
Rb & -0.025 & 0.154 & 0.174 & 0.303 \\ 
Cs & -0.026 & 0.183 & 0.239 & 0.395 \\ 
Fr & -0.035 & 0.221 & 0.450 & 0.636 \\ 
\end{tabular}
\end{center}
\end{table}

\begin{table}
\caption{ Comparison of theoretical and experimental values for Cs $6s-7s$
reduced magnetic-dipole matrix element in atomic units, multiplied by a factor $10^{5}$. The experimental errors are given in parentheses.}
\label{TabExp-Th}
\begin{center}
\begin{tabular}{lr@{.}l@{}l}
\multicolumn{1}{c}{Reference} & \multicolumn{3}{c}{$\langle 6s|| M_1||7s\rangle \times 10^{5} $} \\ \hline 
\multicolumn{3}{c}{Theory} \\ 
This work                                            &  -8&78 &  \\ 
Dzuba \emph{et al.}~\cite{Dzuba85}, 1985       & -13&7 &  \\[0.3pc] 
\multicolumn{3}{c}{Experiment} \\ 
Bennett and Wieman~\cite{BennettWieman99}, 1999   & -10&40 & (0.03) \\ 
Bouchiat and Gu\'ena~\cite{Bouchiat88}, 1988\tablenotemark[1]   & -10&5 & (0.1) 
\end{tabular}
\end{center}
\tablenotetext[1]{ The average of previous experimental results corrected 
for the electric-quadrupole contribution.}
\end{table}

\end{document}